# Filling the Gap between Business Process Modeling and Behavior Driven Development

Rogerio Atem de Carvalho
Rodrigo Soares Manhães
Fernando Luis de Carvalho e Silva

Nucleo de Pesquisa em Sistemas de Informação (NSI), Instituto Federal Fluminense (IFF), Brazil
{ratem, rmanhaes, fernando.carvalho@iff.edu.br}

## 1. Introduction

Behavior Driven Development (NORTH, 2006) is a specification technique that is growing in acceptance in the Agile methods communities. BDD allows to securely verify that all functional requirements were treated properly by source code, by connecting the textual description of these requirements to tests.

On the other side, the Enterprise Information Systems (EIS) researchers and practitioners defends the use of Business Process Modeling (BPM) to, before defining any part of the system, perform the modeling of the system's underlying business process. Therefore, it can be stated that, in the case of EIS, functional requirements are obtained by identifying Use Cases from the business process models.

The aim of this paper is, in a narrower perspective, to propose the use of Finite State Machines (FSM) to model business process and then connect them to the BDD machinery, thus driving better quality for EIS. In a broader perspective, this article aims to provoke a discussion on the mapping of the various BPM notations, since there isn't a real standard for business process modeling (Moller et al., 2007), to BDD.

Firstly a historical perspective of the evolution of previous proposals from which this one emerged will be presented, and then the reasons to change from Model Driven Development (MDD) to BDD will be presented also in a historical perspective. Finally the proposal of using FSM, specifically by using UML Statechart diagrams, will be presented, followed by some conclusions.

## 2. History

This paper represents the evolution of the method initiated in Carvalho (2005), which proposes a way of documenting the connection of a state-based workflow with the entities that realizes it. In parallel, Petrucci et al. (2006), presents an approach that aims to automatically generating formal specifications for state-based workflows using Process Algebra, which in turn allows the formal verification of this workflow using Model Checking. In that way, it is possible to automatically indentify inconsistencies in complex workflows that represent the integration of collections of inter-related business process.

Carvalho and Campos (2006) propose a MDD process for the ERP5 system (Smets-Solanes and Carvalho, 2003), based on the GERAM (IFIP, 1999) framework and a series of modeling and implementation workflows, which in the end culminate into an application of the WARC artifact, proposed in Carvalho (op. cit.). This paper also presents the tools used to support the proposed process, including ERP5 Generator, a code generation tool capable of generating structure, basic behavior and GUI for a new or extended system module. Other tools, for supporting project management are also presented briefly.

Further, Carvalho and Campos (2007) presents the Quality Assurance techniques of the proposed ERP5 development process. At this point, it was considered that "code generation avoids common programming mistakes, reducing testing activities to the code manually written during the code completion activity". For the manually written code, a tool called ERP5 Test Case was used to provide template testing scripts that automate most of Unit and Integration tests, and Zelenium, a Zope GUI test tool provides user interface testing.

Finally, Monnerat, Carvalho and Campos (2008) presents the last evolution of the proposed

system development process started in Carvalho (2005), in this case, again, specific for the ERP5 system, and pointing towards the creation of Deployér, a tool to support all the development and customization process for ERP5.

**3. The Turning Point**

Up to Monnerat et al. (2008) the focus of the process was in MDD, in fact, Khajeh-Hosseini, Sommerville and Sriram (2010) state that Monnerat et al. (op. cit.) represents relevant background for modeling applications and application portfolios in the arena of enterprise systems using MDD. However, MDD is not a panacea, and Agile techniques are proving their efficiency and financial value. Following this path, Carvalho, Johansson and Manhaes (2009) and Carvalho, Johansson and Manhaes (2010) supply mappings (and limitations) of Agile Methods to enterprise information systems development and customization.

At the same point, this team of researchers and developers embraces Behavior Driven Development, creating in 2009 Pyramid[1], a stack to provide all tools necessary for the use of BDD in Python.

**4. BDD and Business Process Modeling**

However, there is still a gap between BDD and BPM, and this gap is the fact that while BDD uses textual descriptions of the business process, the enterprise information systems community is used to develop graphical representations of the business process, based in UML diagrams for instance. The question was so to create a kind of business process-focused BDD, by substituting the textual description of requirements to some graphical notation. At this point, Martin (2008) has already pointed that the triad GIVEN-THEN-WHEN of BDD is nothing more than the description of state transitions and therefore, BDD scenarios could be represented by Finite State Machines (FSM) and vice-versa, with all advantages that this could bring, such as a plenty of tools available for Formal Verification, as stated in Carvalho e Campos (2006).

The answer to the problem of connecting BDD to BPM is to model business process using FSM and then connect these representations to the BDD stack, or, in terms of programming, to substitute the parsing of text by parsing of XMI representations of UML Statechart Diagrams. Moreover, connecting BDD tools to the process originally proposed by Carvalho e Campos (op. cit.) would provide an even higher level of automation of the customization and development of ERP modules.

Although FSM are less used in BPM, being Petri Nets based notations growing in acceptance, according to Aalst and Hee (2004), "it is, most of times, straightforward to make translations between Statechart Diagrams and Petri Nets", allowing the translation of business processes modeled in Petri Nets to FSM and them following the BDD path, or even providing a tool that maps directly from Petri Nets to the GIVEN-WHEN-THEN structure.

**5. The Proposal[2]**

The proposal here presented is to substitute the textual requirements descriptions of BDD to a Statechart Diagram. Also, we propose to use a Domain Specific Language (DSL) to describe the conditionals and other textual elements of the FSM. In that way, it is expected that a user can define the business process through a Statechart Diagram and using an appropriate language – instead of the underlying programming language used to implement the system. In fact, given a diagram in which the textual elements are written using a DSL, it can be converted into any language, given that there is a proper stack of BDD tools.

Moreover, it is possible, by joining a proper GUI testing tool, to provide "live" demonstrations of the business process realized by a given information system, allowing the user to visually checking the execution of a given workflow.

---

1 http://www.renapi.org/biblioteca-digital/ferramentas
2 This proposal fits not only for ERP5 but to any system in any language, provinding that there is a BDD stack available.

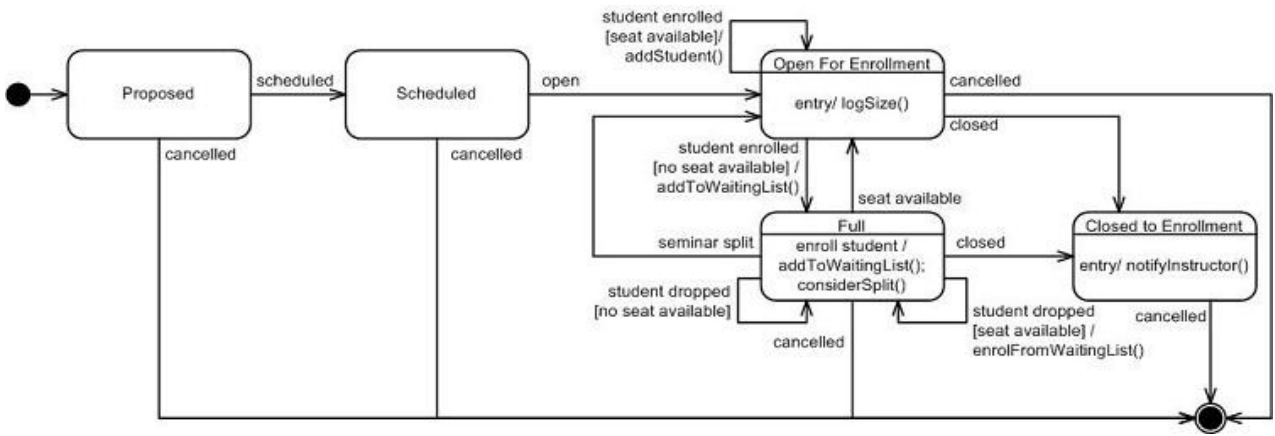

Figure 1: An example statechart representing a business process (Source: Ambler, 2009)

In order to illustrate this proposal, it is used an example obtained from Ambler (2009): "The arrows in Figure 1 represent transitions, progressions from one state to another. For example, when a seminar is in the Scheduled state, it can either be opened for enrollment or cancelled.  The notation for the labels on transitions is in the format event [guard][/method list].  It is mandatory to indicate the event which causes the transition, such as open or cancelled.  Guard, conditions that must be true for the transition to be triggered, are optionally indicated. The [not seat available] guard is shown on the student enrolled transition from the Open For Enrollment to the Closed To Enrollment state." Ambler (op. cit.) also states that "Guards can be described in any manner, including both free form text or formal language – when I'm whiteboarding I'll use free form text to ensure that it's readable by everyone but with a sophisticated CASE tool I would consider using either a programming language such as Java or a modeling language such as Object Constraint Language (OCL) if the tool has the ability to actually process that information into something useful (such as executable code)", in the BDD case the Domain Specific Language must be used, for instance by defining expectations that simplify a lot the business process description by the client/user.  In that way, we would go back to the whiteboard by using (almost) free texting – making the diagram in Figure 1 more readable for non-technicians.

The invocation of methods, such as addToWaitingList() are optional, but since we are in the requirements level, they shouldn't be represented by the diagram – since they provide useful information to the user. In BDD parlance the states Scheduled, Finished and Proposed in Figure 1 could be translated in the following way:

GIVEN I have the proper rights
WHEN I open a new course
THEN the course is officially p**roposed**

GIVEN the course is **Proposed**
WHEN I define its schedule its starting date as <mm1/dd1/yyyy1> and its finish date as <mm2/dd2/yyyy2>
THEN it should be **scheduled** to occur between <mm1/dd1/yyyy1> and  <mm2/dd2/yyyy2>

GIVEN the course is **Scheduled**
WHEN I cancel it
THEN the process is **finished**

GIVEN the course is **Scheduled**
WHEN I open it to enrolment with <logsize>

THEN the course should have a <logsize> number of seats available

GIVEN the course is open for enrolment AND there is a seat available
WHEN one student enrolls
THEN the student is associated to the course AND the number of seats available is decreased of one

Again, it is important to note that with a DSL properly defined it is possible to describe the statechart in Figure 1 using a language closer to the user jargon. Another matter is to consider or not all the possibilities of UML Statechart Diagrams, it is necessary to balance business process representation richness with users' understanding. In most cases the information in the transitions given by the notation "event [guard condition]/ action" is sufficient, in some others entry and exit actions would also help. Further research must determine how to map nested and parallel states to BDD.

## 6. Conclusions

The intention in using FSM is not to substitute BDD's GIVEN-WHEN-THEN triad but instead to give users the option of also using graphics for representing business processes. If the user brings a business process model, the FSM should be used, instead, if he/she prefers to describes the process in plain text the "traditional" BDD is used. Moreover, translations can be provided from other notations such as UML Activities Diagrams or "pure" Petri Nets representations. The user of graphical notations has the advantage of adding visual information to the model, which in turns can provide better ways of re-using parts of business processes and even providing "live" presentations of the processes by running the automated tests. Therefore, this proposal opens the path for the creation of a series of tools:

- The basic one associated to this proposal, which will translate from a XMI representation of a UML Statechart Diagram to a set of @GIVEN-@WHEN-@THEN statements, for the first level of connection requirements-source code;
- One capable of, giving a graphical representation of the statechart, running the system step by step, allowing the user to choose the path to follow in the diagram and executing visually the system using the automated tests' examples as inputs;
- A translator capable of, given a statechart diagram, creating the correspondent text, and vice-versa, by creating a XMI from a text.
- The same set of tools, but using Petri Nets as the modeling paradigm.

In that way it would be possible to reduce a series of problems related to the translation from business process models to source code, reducing customization and implementation times and costs for different kinds of Enterprise Information Systems.